# Estimation of Playable Piano Fingering by Pitch-difference Fingering Matching Model

Haoyue Zhao[a], Xin Guan[a], Qiang Li[a]

[a]School of Microelectronics, Tianjin University, Tianjin 300072, China

**Abstract**

The existing piano fingering labeling statistical models usually consider the constraints among the fingers and the correlation between fingering and notes, and rarely include the relationship among the notes directly. The limited learned finger-transfer rules often cause that some parts of the fingering cannot be playable in fact. And traditional models often adopt the original notes, which cannot help to explore the mapping nature between the pitches and fingering. Inspired from manually annotation which acquire the fingering knowledge directly from pitch-difference, we proposed a pitch-difference sequence and fingering (PdF) matching model. And to get playable fingering, besides learned finger-transfer rules, prior finger-transfer knowledge is especially combined into the model. In order to characterize the playable performance of the model, we also presented a new evaluation index named incapable-performing fingering rate (IFR). Comprehensive experimental results show that compared with the existing state-of-the-art third-order hidden Markov labeling model, the general and the highest matching rate of our model increases by 3% and 1.6% respectively, and the fingering for all scores can be playable.

## 1 Introduction

The keyboard music performance is affected mostly by hand related movement, especially the choice of finger. Though finding proper fingering for the keyboard instruments is one of the necessary skills, it is non-trivial for novice, even for virtuoso. Traditional fingering annotation depends on trial and error based on experience which increases the difficulty for beginners and amateurs. While from optimization viewpoint, fingering estimation for keyboard instruments is a combinatorial problem in nature and can be done automatically. Automatic fingering estimation has been an emerging topic of musical symbol information processing. And it has been promoted machine understanding of musical symbol information, and strengthened instrument practice assistant [1], education systems [2] and music content arrangement [3].

At present, automatic fingering estimation strategies is mainly divided into rule-based models and data-driven ones. Rule-based methods take anatomic constraints and performance difficulty of finger pairs as cost of the finger sequences to predict optimal fingering [1,4-8]. And the problem lies on the lack of theoretical and empirical evidence for determining parameters. Data-driven schemes address this limitation by learning the parameters from note sequences and corresponding fingering ones with statistical model. In turn, the quality and quantity of the data set is one of important factors to affect the predict performance which is vital for deep learning method. Besides the data, predict performance also depends on to what degree the model or network reflects the relationship between the musical context and finger sequence and combines constraints and prior knowledge of fingering. To the best of our knowledge, statistical learning method with sufficient size of data outperform the rule-based one according to the annotation accuracies.

The optimal fingering strategy in any situation rarely exists because some emphasize on precision and speed, some phrasing and dynamic articulation, some memorization, others musical interpretation, not to mention difference of hand size and shape. In our data-driven method, fingering indicator are mainly from the sheet music score and experienced pianist [9].

The fact that finger choice is mainly related to interval between neighboring pitch(s) pair than original note sequence, inspire us to model the connection between pitch-difference sequence and fingering (PdF), instead of original note sequence and fingering (NF). This strategy can decrease the variation of the musical symbol sequence, which in turn reduce model complexity, alleviate the demand of data size, and improve the annotating accuracy. Besides this, finger transfer is considered when note sequence is ascending or descending without hand movement. Though the basic bidirectional long short-term memory (BI-LSTM) network can include long-range pitch interval information and the relationship between the pitch interval and the fingering, it cannot reflect the connection between finger pairs directly. These observations shed lights on refining the network of BI-LSTM to combine prior knowledge between finger sequences and musical symbol context. When the model predicts the finger sequence, a prior finger transfer rules is introduced, which can eliminate impractical fingering.

To improve generalization performance limited by small data sets, an augmented set based on the statistics of PdF in the existing data set is constructed. When evaluating the fingering, we also stress on the playability of the fingering to ensure it can work in practice and proposed a measure incapable-performing fingering rate (IFR). The comprehensive experiment results show that the refined model can effectively improve the consistency rate of the fingering annotation.

Our main results are as follows:

1. Presenting a pitch-difference fingering (PdF) model. Compared with the original note sequence, the pitch difference can more intuitively reflect the player's definite fingering habits of the score.

2. Proposing new method for expanding dataset. By counting the frequency and transfer rules of fingerings in existing data sets, an augmented set is generated and added to the network training stage.

3. Introduce a prior fingering knowledge, and correct the fingering relationship within an octave that should not appear.

4. The addition of new fingering results evaluation measure, which can better assess the quality of fingering.

We review current piano fingering annotation methods in Section 2. Method for piano fingering estimation in Section3, and evaluation of results are presented in Section4. We conclude in Section 5.

**2 Related Works**

Since the first article on fingering estimation was published in 1997, the current research on fingering estimation is mainly divided into methods based on rules or costs and methods based on data-driven.

Parncutt et al. [4] designed the first calculation model of piano fingering. This model was built for the finite-length note segment played by the right hand. Considering the comfortable span between the fingers and the span between the notes, they established the rules based on the common fingering logic of pianists. For long sequences, the authors used rules to define the cost function of fingering costs, and used traditional dynamic programming methods to find the path with the lowest fingering cost. According to the characteristics of different works, this method required manual adjustment of rule weights to achieve better results. Jacobs et al. [5] later improved this model. They replaced the semitone measurement in the original model with physical distance to compensate for the inappropriate difficulty ranking caused by semitone errors. Nellåker E et al. [6] introduced an additional rule of pause based on the 12 rules established by Parncutt et al. to evaluate the difficulty of fingering before and after the pause. Lin CC et al. [1] designed a Sliced Fingering Generation (SFG) fingering generation algorithm, which split the score into segments with cross content. And they used the rules suggested by Parncutt et al. [4] to define the cost function, finally used dynamic programming to generatereal time Piano fingering. Hart et al. [7] described a dynamic programming method for the right-hand segment as a state transition constraint. Al Kasim et al. [8] defined the horizontal cost of adjacent notes and the vertical cost of chords, and used the grid graph generated by music fragments to find the path. This model expanded the fingering recognition of polyphony in fingering recognition. Examples have proved the effectiveness of the model, but the model lacks quantitative evaluation criteria and has low constraint compatibility.

Introducing rules or costs is indeed in line with the process of determining fingerings in terms of logical understanding, but when the rule model is used for more musical scores to obtain fingerings, it is often necessary to modify the internal parameters of the model again.

Yonebayashi et al. [10] proposed the first recognition method of piano fingering based on statistical modeling of the corresponding characteristics of pitch and fingering. They used Hidden Markov Model (HMM) to model the fingering sequence. The finger number corresponding to the note is the fingering state to be predicted, and the probability of each fingering state depends on the fingering state at the previous time and the note output at the current time [10]. However, the assumption of independent observation leads to the fact that the fingering annotations do not contain the information of adjacent pitches, and the information of adjacent pitches cannot be used to further restrict the probability of fingering transition. Later, Nakamura E et al. [11] proposed a "merged HMM" to automatically separate and label the unseparated left-hand and right-hand fractions. Li Qiang et al. [12] combined the prior knowledge of fingering rules, introduced the judgment function, and improved the optimization rules of the Viterbi algorithm.

The advantage of the statistical model is that it can optimize the parameters based on the statistical learning from the data. But it only pays attention to the most local fingering constraints of continuous notes, so that part of the effective information, such as the long-range fingering relationship, is ignored. And because of the improvement of model adaptability, the error rate of the results is higher when some special fingerings appear. If you want to increase the Hidden Markov Order to fit the optimization results of the fingering long-range relationship, the experiment shows that Under the small data set, the third-order Hidden Markov model has no obvious improvement. [9]

Nakamura et al. [9] conducted a statistical analysis on the individual differences of piano fingering in their review article. In addition to the statistical model, the fingering labeling with a deep neural network (DNN) model through labeled data was mentioned. The authors took the pitch as input and the corresponding number of the fingering as output, used the feedforward (FF) network and the long short-term memory (LSTM) network to label the piano fingerings. The results are slightly lower than the statistical model results of the same data set. The FF network and the LSTM network do not have constraints between the output layer units, so that the model ignores the dependency between fingerings, and the data does not distinguish between monophonic and polyphonic notes, which leads to an increase in finger reuse rate and non-finger fingering rate of chords.

In this article, the preliminary data processing strengthens the correspondence between notes and fingerings. BI-LSTM network analyzes the contextual pitch relationship and the corresponding relationship between pitch and fingering. We use the fingering transition matrix that characterizes the fingering relationship to optimize the annotation results. In the labeling processing, the strong restriction based on the fingering rules results in the final fingering labeling sequence. Experiments have verified that besides good recognition of basic fingering, it also has good performance in the special fingering, like turning the finger and using different fingers to play the same key. It has the advantages of both the constraint model and the statistical model. Compared with the existing labeling model, the match rate is improved.

## 3 Methods

Our PdF annotation network has a recursive structure and includes 3 layers, as shown in Fig. 1. The first layer is note converting to represent pitch-difference. The middle layers constitute a BI-LSTM like network to capture the forward and backward context information of the pitch-difference in the music score, and investigate the probabilistic relationship between the pitch-difference and the fingering. The third layer is to integrate the finger transfer regulations. Particularly after training, the finger transfer matrix from established fingering rules is added that can trim the impossible fingering path, avoid labelling adjacent pitches of the same finger, and ensure the playability of the predicted fingering.

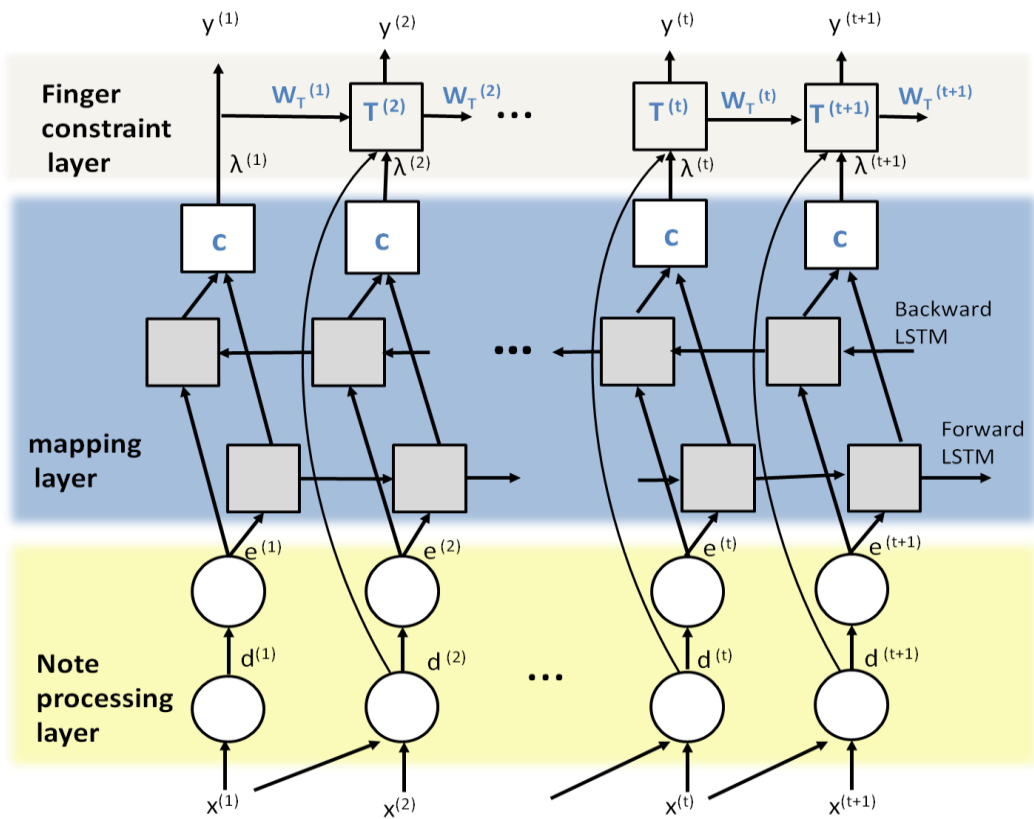

Fig. 1 Pitch-difference Fingering annotation network

### 3.1 Pitch-difference representation method discriminating monophony from polyphony

Original notes are converted to pitch-difference at first, because, when annotating fingering, the point that human mainly focus on is the change of the neighbor pitches, other than every concrete pitch. As to pitch-difference, though traditional concept of interval measures pitch distance between two tones, it is not discriminated between monophonic and polyphonic situations, and cannot be used directly in our model. In fact, pitch-difference in melodic or harmonic dimension affects the choice of fingers to a great extent, so we adopt distinct ways to express the same pitch-difference in melody and in harmony.

We extracted the note ID, pitch start time, offset time, hand information and fingering from the original data. After separating the left-hand data from right-hand, we combine the pitch start time and offset time, and convert the note ID to MIDI number. Then, expand the notes in chronological order into a sequence, and for harmonic notes, the order is from low pitch to high one. At last, calculate the pitch-difference $d^{(t)}$ according to Formula (1). A converting example is shown in Fig. 2.

$$d^{(t)} = \begin{cases} n \cdot 100 & t = 1 \\ x^{(t)} - x^{(t-1)} + n \cdot 100 & |x^{(t)} - x^{(t-1)}| < 12, t > 1 \\ 80 sgn(x^{(t)} - x^{(t-1)}) & |x^{(t)} - x^{(t-1)}| \geq 12, t > 1 \end{cases} \quad (1)$$

Where n represents number of notes with same onset. And for single tone, n equals 0.

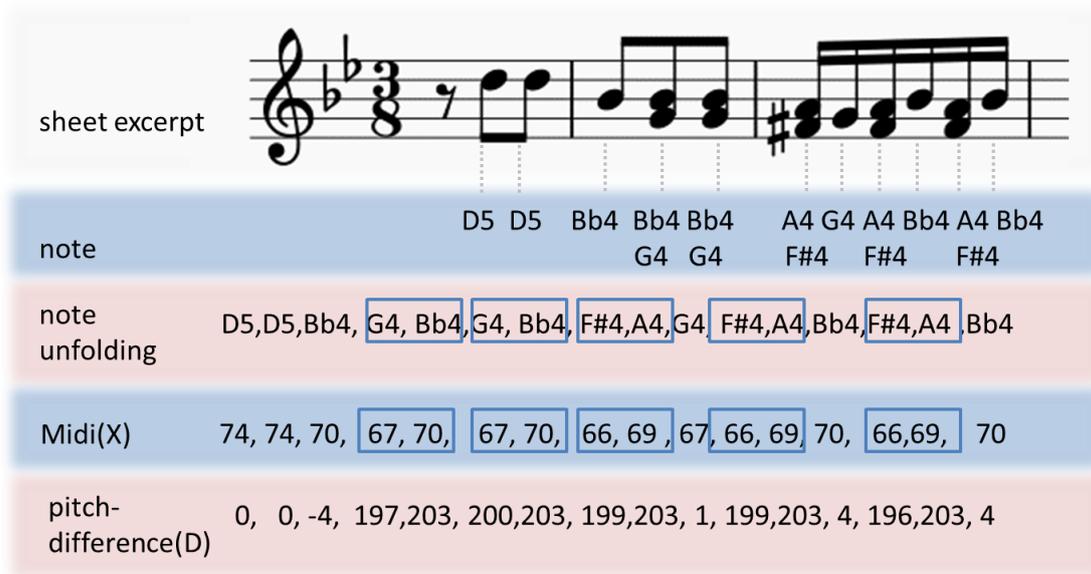

**Fig. 2** Calculate procedure of pitch-difference sequence of a score excerpt

Since the total size of pitch-difference sequence in the actual conversion is 1 less than the fingering. In order to realize the one-to-one correspondence between the pitch-difference sequence and the fingering, the first position of the sequence is supplemented as shown in the case of t=1. When t is 1, if the initial sound is a single tone, adds 0. If it is a polyphony, multiply it by 100 according to the number of notes contained in the polyphony. The number "100" is chosen to separate polyphonic notes of different number.

When the current pitch-difference is greater than one octave, that is, when the Midi difference is greater than or equal to 12, the fingering is relatively simple. In this situation, $d^{(t)}$ only represents the ascending or descending of the scale, other than specific numerical difference of the physical position, which is uniformly expressed as 80. Before input into our model, $d^{(t)}$ is represented by integer and encoded into a vector $e^{(t)}$.

### 3.2 Pitch-difference finger relationship mapping layers

The middle layers constitute a BI-LSTM like network, where two LSTM layers capture the forward and backward context information of the pitch-difference in the music score, and one mod-els the probabilistic relationship between the pitch-difference and the fingering. In the forward and backward LSTM network, the basic unit is LSTM cell whose internal gates are designed to obtain long-range relationship of pitch-difference to make good use of the context to decide the fingering. The long-range context is especially important on situations, such as learning cross fingering in ascending or descending scale, or changing fingers in order for the fast repeated notes.

**3.2.1 Pitch-difference fingering relationship feature mapping layer**

PdF relationship feature mapping layer is the BI-LSTM layer in our model. Forward and backward LSTM are used to mine the note finger relationship. For certain note, fingering annotation depend on the notes before and after, so we adopt bi-directional network to find the relative position of fingers.

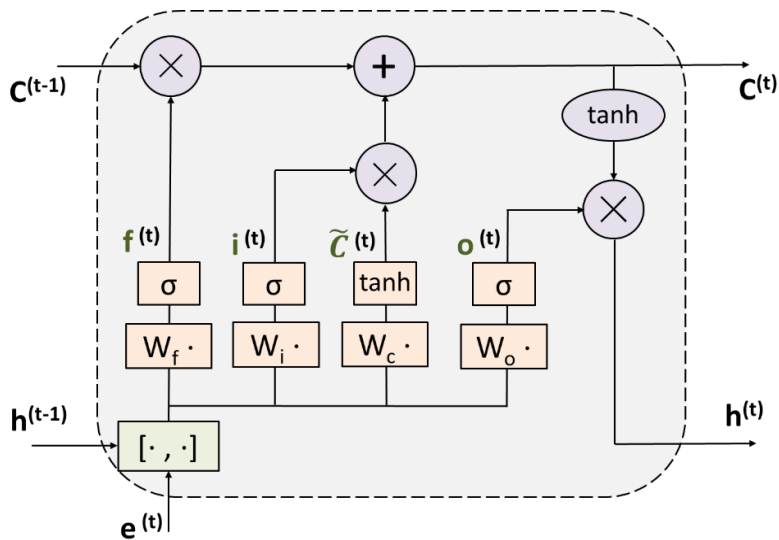

**Fig. 3** A structure of LSTM cell

For the basic unit of the LSTM, there are three gates to realize its function, forget gate, input gate and output gate[16]. At a certain instant t, the detailed forward process of the training and the components are updated as follows:

The forget gate discards the useless long-term memory and judges whether to continue to store the long-term score information. For example, the previous chord input has little effect on the current fingering, and the forget gate can discard the long-range score information.

$$f^{(t)} = \delta(W_f \cdot [h^{(t-1)}, e^{(t)}] + b_f) \tag{2}$$

where $W_f$ is the weight matrix, $W_f \in R_{hidden_{size} \times (hidden_{size} + e_{size})}$, $[h^{t-1}, e^t]$ is the concatenation of two vectors, the input pitch-difference information $e_t$ and the former hidden state $h_{t-1}$. $b_f$ is the bias of the forget, and $\sigma$ is the sigmoid function $\sigma(z) = \frac{e1}{1+e^{-z}}$.

The input gate controls whether put current input into cell state.

$$i^{(t)} = \delta(W_i \cdot [h^{(t-1)}, e^{(t)}] + b_i) \tag{3}$$

$$\tilde{C}^{(t)} = tanh(W_c \cdot [h^{(t-1)}, e^{(t)}] + b_c) \tag{4}$$

Where $W_i$ is the weight matrix, $b_i$ is the bias, and $tanh(z) = \frac{e^z - e^{-z}}{e^z + e^{-z}}$.

Cell current state $C^{(t)}$ is a combination of previous memory and current memory. Under the control of the forget gate, it can save the music score information from a long time ago. And under the control of the input gate, it can prevent the current irrelevant content from entering the memory.

$$C^{(t)} = f^{(t)} * C^{(t-1)} + i^{(t)} * \tilde{C}^{(t)} \tag{5}$$

Where $*$ means multiply by element. This calculation will input the real-time score information state into the long-term memory.

The output gate controls the impact of long-term memory on the current output.

$$o^{(t)} = \delta(W_o \cdot [h^{(t-1)}, e^{(t)}] + b_o) \tag{6}$$

Where $W_o$ is the weight matrix, and $b_o$ means the bias.

The state of the hidden layer is updated to control whether the long-term state $C^t$ is output as the finger label of the current LSTM cell.

$$h^{(t)} = o^{(t)} * tanh(C^{(t)}) \tag{7}$$

The final output of the LSTM cell is determined by the output gate and the cell state. If the hidden layer state is extracted at each time step, we can get the output sequence with the same size as input. So, the LSTM network can theoretically model the correspondence between the note sequence and the fingering. As to BI-LSTM, it is a combination of forward LSTM and backward LSTM. And the output is the concatenation of the front and back hidden vectors $[h^{(t)}{}_{(forward)}, h^{(t)}{}_{(backward)}]$.

### 3.2.2 Probability-fingering mapping layer

In order to better characterize the fingering recognition results, the hidden state vector of $2 * hidden\_embeding\_size$ is mapped to k-dimension in the connectivity layer of BI-LSTM backend, where k is the number of fingering labels and $\lambda^{(t)}$ represents the probability of each fingering label for the current input.

$$\lambda^{(t)} = c[h^{(t)}{}_{(forward)}, h^{(t)}{}_{(backward)}] \tag{8}$$

$$\lambda^{(t)} = (\lambda_1^{(t)} + \lambda_2^{(t)} + \lambda_3^{(t)} \cdots + \lambda_k^{(t)})^{\mathrm{T}} \tag{9}$$

$\lambda_j^{(t)}$ indicates that the BI-LSTM model is at the t-th position of the score, and the probability of the fingering label with index j.

### 3.3 Finger transferring layer

The finger transferring layer is added to the output of the BI-LSTM. Though the BI-LSTM model can include long-range context information in the note sequence, and learn the correspondence between music notes and fingering, it cannot represent the ergonomic constraints between fingers. In order to address this problem, we introduce the finger transition probability matrix $W_T$ into the model to restrict two adjacent fingerings in the output of BI-LSTM. The finger transition is related to the ascending and descending of the neighbor monophony, so the model introduces different fingering transfer matrices to the BI-LSTM output according to the input type.

$$W_T = \frac{1}{2}[sgn(d^{(t)}) \cdot (W_{T\uparrow} - W_{T\downarrow}) - sgn(d^{(t)}) \cdot (W_{T\uparrow} + W_{T\downarrow})] \qquad (10)$$

Where $W_{T\uparrow}$ is finger transition probability matrix of ascending pitches, and $W_{T\downarrow}$ is of descending pitches. In particular, $W_{T\uparrow}$ and $W_{T\downarrow}$ only restrict the fingering of the adjacent single tone. The constraint function of matrix $W_T$ to finger is shown in Fig. 4. Where $P_{ij}$ is the transfer possibility of finger i to j one.

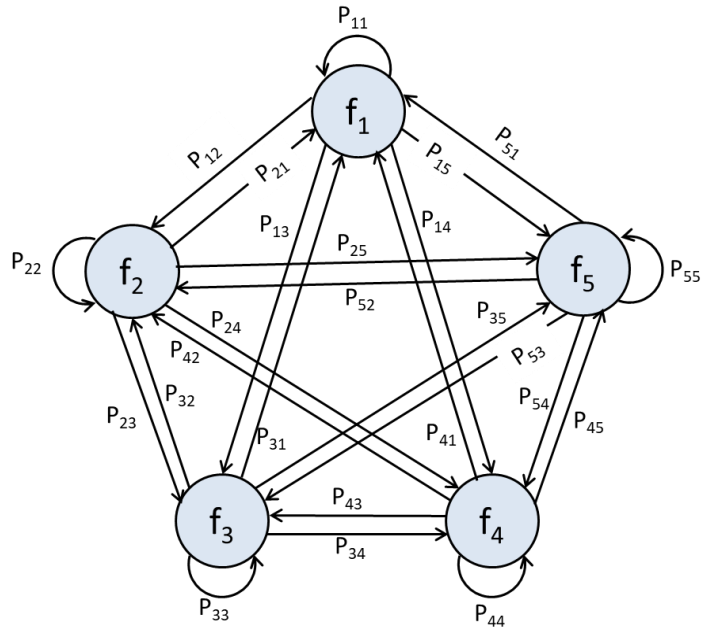

**Fig. 4** Schematic diagram of fingering transfer

For input pitch-difference sequence D, the fingering transfer matrix $W_T$, and the output fingering sequence $\Lambda$ of BI-LSTM to the score information, the output Y is expressed as follows:

$$\widetilde{y^{(t)}} = W_T * y^{(t-1)} + \lambda^{(t)} \qquad (11)$$

$$\hat{y}^{(t)} = softmax(\widetilde{y^{(t)}}) \qquad (12)$$

$$Y = (\hat{y}^{(1)}, \hat{y}^{(2)}, \cdots \hat{y}^{(n)}) \qquad (13)$$

And $y_i^{(1)} = \lambda_i^{(1)} = \Lambda\,(1,i)$, $y^{(t)}$ can also be expressed as the following form:

$$y^{(t)} = (y_1^{(t)} + y_2^{(t)} + y_3^{(t)} \cdots + y_k^{(t)})^T \qquad (14)$$

$y_i^{(t)}$ represents the result of the occurrence probability of the fingering label with index i in the t-th position of the score. Then at time t, the fingering with the highest probability, that is, the model's fingering $\varphi^{(t)}$ for the current pitch is:

$$\varphi^{(t)} = \arg\max_i \left[ y_i^{(t)} \right] \quad (15)$$

**3.4 Prior finger transfer knowledge constraint**

Through we optimize finger transition matrix during training, it can only learn weak relationship of neighbor fingers subjected to model structure and data variety. Assuming that fingers other than the thumb do not cross over each other, the kinematic characteristic of the left-hand and right-hand fingers makes some fingerings definitely impossible if we discriminate hand and movement direction in detail, as shown in table 1. Take the right hand as an example, when the scale is descending, the transition from finger 2, 3, or 4 to finger 1 is the possible, and from finger 2 to finger 3 is impossible without hand movement.

**Tabel.1** Possible (√) and impossible (×) finger transition for Left-hand ascending scale/right-hand descending scale without hand movement

| $f^{(t-1)}$ \ $f^{(t)}$ | 1 | 2 | 3 | 4 | 5 |
|---|---|---|---|---|---|
| 1 | √ | √ | √ | √ | × |
| 2 | √ | √ | × | × | × |
| 3 | √ | √ | √ | × | × |
| 4 | √ | √ | √ | √ | × |
| 5 | √ | √ | √ | √ | √ |

The possible and impossible finger transition for left-hand ascending scale and right-hand descending scale when playing monophonic note pair without hand movement is summarized in table 1. The left-hand descending scale and the right-hand ascending scale are diagonally mirror-symmetrical with Table 1. In predicting period, we use this prior knowledge to ensure the playability of the fingering.

The decision function $T$ is established according to Table 1. If the current fingering is an impossible one, the transition probability between fingers is set to 0, otherwise keep the current output of the model. In the ground truth of left fingering, there is a 5th finger to 1st finger hand movement within an octave, so prior knowledge does not restrict this fingering conversion.

$$T = \begin{cases} \frac{1}{2}\{sgn[4.5 - f^{(t-1)} \cdot f^{(t)}] + 1\} & right: -12 < d^{(t)} < 12 \text{ and } sgn(d^{(t)}) \cdot (f^{(t)} - f^{(t-1)}) < 0 \\ \frac{1}{2}\{sgn[5.5 - f^{(t-1)} \cdot f^{(t)}] + 1\} & left: -12 < d^{(t)} < 12 \text{ and } sgn(d^{(t)}) \cdot (f^{(t)} - f^{(t-1)}) > 0 \\ 1 & other \end{cases} \quad (16)$$

When the decision function equal 0, there will be fewer fingering transfer paths between the two moments. As shown in Fig. 5, in left-hand ascending scale or right-hand descending scale, some transfer paths are trimmed.

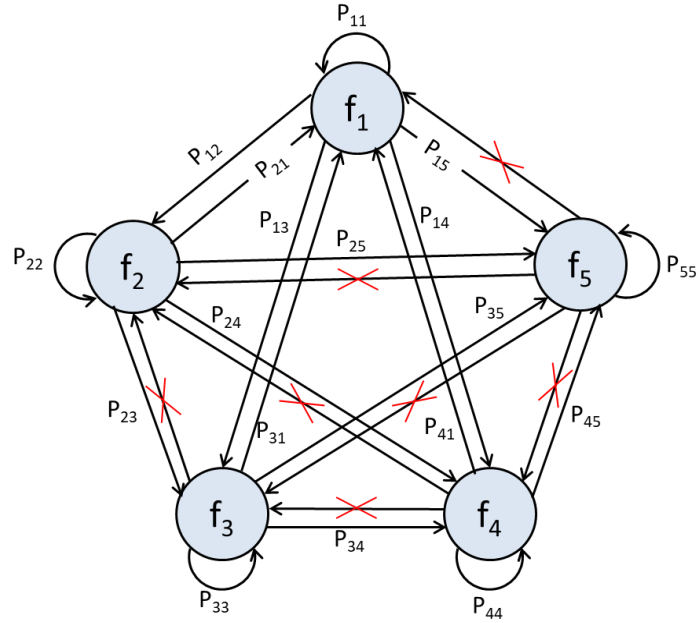

**Fig. 5** Schematic diagram path search when state is limited

After adding the path pruning condition, the expression of $\widetilde{y^{(t)}}$ in the recognition period is as follows.

$$\widetilde{y^{(t)}} = (T \cdot W_T)^T * y^{(t-1)} + \lambda^{(t)} \quad (17)$$

At time t, we choose the fingering with the highest probability as the final fingering estimation, the model's fingering $\varphi^{(t)}$ for the current pitch is:

$$\varphi^{(t)} = \arg\max_{i} \left[ \widetilde{y_i^{(t)}} \right] \tag{18}$$

## 4 Evaluation

### 4.1 Setup

The experiment data comes from the first public accessed dataset- Piano Fingering(PIG) Dataset[9] and an augmented set generated via the piano fingering rules. The PIG dataset has 150 pieces scores and 309 music fingering data. Due to one piece of score may correspond to multiple fingering, the dataset is split into 35 non-overlapping subsets. And fingerings belonging to the same score are all in the same subset. In ablation study, we use 35-fold cross-validation to train and test the model. And we used the Adam optimizer for training [17]. In the 4.3.1 accuracy comparative experiment, we used a three-layer LSTM network. And in order to reflect the comparability of the results, in the 4.3.6 comparative experiment we use the same testing set as [9]. The method of generating the additional augmented set added to the training set is described in appendix. And if there is no additional description below, all the results marked with "our model" are the results obtained by adding augmented set for training and using the model shown in Fig. 1.

### 4.2 Evaluation measures

#### 4.2.1 Matching rate

In theory, the higher the coincidence rate between the labeling result and the label, the better the algorithm effect. Suppose $L_h$, $L_a$ and $R_h$, $R_a$ are the fingering sequences of left- and right-hand manual annotation and algorithm annotation respectively, the music score total length is n, and the calculation formula of matching rate is as follows:

$$\alpha = 1 - |L_h \wedge L_a + R_h \wedge R_a| / n \tag{19}$$

Refer to the evaluation method of the paper [9] to compare the effectiveness of our model. We use evaluation measures that can be used for multiple ground truth data. When there are multiple basic facts, calculate the matching rate between the estimated value and each basic fact and find the average value, expressed as the general matching rate $M_{gen}$. And based on the closest basic fact of the estimated value defines the highest matching rate $M_{high}$. The score data size is N, and after adding different fingering tags, the training data is extended to $N_{gen}$. The general matching rate and the highest matching rate calculation formula are as follows:

$$M_{gen} = \sum_{i,j} \alpha_{ij} / N_{gen} \tag{20}$$

$$M_{high} = \frac{\sum_i \max_j \alpha_{i,j}}{N} \tag{21}$$

**4.2.2 Incapable-playing fingering rate**

The physically playable fingering sequences. Therefore, the percentage of "unachievable fingering" is called incapable-performing fingering rate (IFR). The calculation formula is as follows.

$$IFR = \sum_i \frac{\sum_{t=2}^{n_i} \psi(d^{(t)}, f^{(t)}, f^{(t-1)})}{ni} - s / N \tag{22}$$

$$\psi(d^{(t)}, f^{(t)}, f^{(t-1)}) =$$

$$\begin{cases} \frac{1}{2}\{sgn[f^{(t-1)} \cdot f^{(t)} - 4.5] + 1\} & \begin{array}{l} left: -12 < d^{(t)} < 12 \text{ and } sgn(d^{(t)}) \cdot (f^{(t)} - f^{(t-1)}) > 0 \\ right: -12 < d^{(t)} < 12 \text{ and } sgn(d^{(t)}) \cdot (f^{(t)} - f^{(t-1)}) < 0 \end{array} \\ 0 & other \end{cases} \tag{23}$$

In the fingering data, there are some hand movements within one octave. At this time, the prior rules are invalid for them, so we will record the number of fingerings that are consistent with the ground truths but marked as wrong by the prior rules as $s$. $\psi$ is set to 1 for all non-finger fingerings that appear. For example, when the pitch of the left hand drops without changing hands, fingers 3 to 2 are not allowed. At this time, $\psi(d^{(t)}, f^{(t)}, f^{(t-1)}) = 1$.

### 4.3 Results and discussion

### 4.3.1 Pitch-difference modeling strategy

As to PIG dataset [9], when pitch-difference sequence is adopted, 555 right-hand monophonic and polyphonic note combinations are reduced to 108 pitch-differences data with time information, and the left-hand is compressed from 564 to 101, which reveals the note-fingering relationship matching model can be optimized further, and for the same data set, PdF strategy is readily to improve the annotation performance than original one. In the case where the matching network is LSTM and BI-LSTM, compared with NF, the $M_{gen}$ of the PdF modeling strategy is increased by 12.76% and 11.33%, $M_{high}$ is increased by 12.98% and 12.46%, and the IFR is reduced by 4.03% and 2.54%. $M_{gen}$ and $M_{high}$ of PdF strategy have increased greatly. Compared with original note, the pitch-difference information is more directly related to fingering, which is consistent with intuition when manually determining fingering.

### 4.3.2 Bidirectional network

We can see the improvement of bidirectional network on $M_{gen}$ and $M_{high}$ in Fig. 6. This is also confirmed by the fingering estimation logic that needs to consider the position of notes before and after the same time in actual fingering estimation. Through their respective IFRs, it can also be seen that considering the context can also reduce the generation of unplayable fingerings, allowing the model to better understand the correspondence between note and fingering.

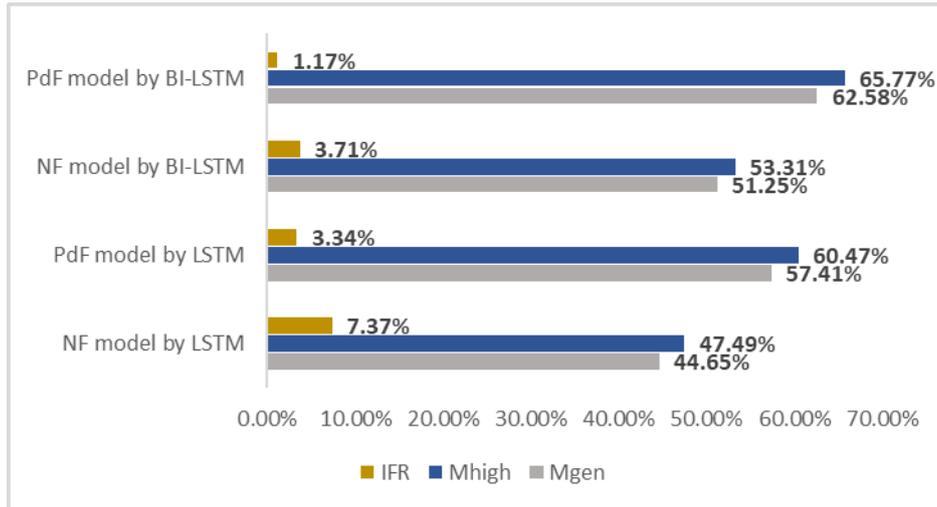

**Fig. 6** Results of ablation study

**4.3.3 Adding learned fingering transfer knowledge**

When evaluate the effect of introduction of finger relationship knowledge to the model, we compare the performance of BI-LSTM with learned transfer constraint and BI-LSTM-CRF both modelling pitch-difference fingering, as shown in C and E of Fig. 7. After adding CRF to baseline model BI-LSTM, performance of the model becomes worse, which means that CRF has not learned the correct finger transfer regulations between fingers.

In BI-LSTM-CRF, conditional random field (CRF) at the backend of BI-LSTM introduces constraints for output, which achieved good results in part-of-speech tagging. [18] It uses the logical relationship between the tags to establish the connection between the output of BI-LSTM network. While as to fingering labeling, it is not enough to only consider transfer statistics on the fingering tags to fit the physiological structure characteristics of hands. From Table 1 and the basic common sense of piano fingering, we can see that the logical relationship of fingering is related to the ascending or descending of the pitch, that is, the type of input. Our fingering transfer layer is theoretically better than adding CRF, and the experimental results also prove this.

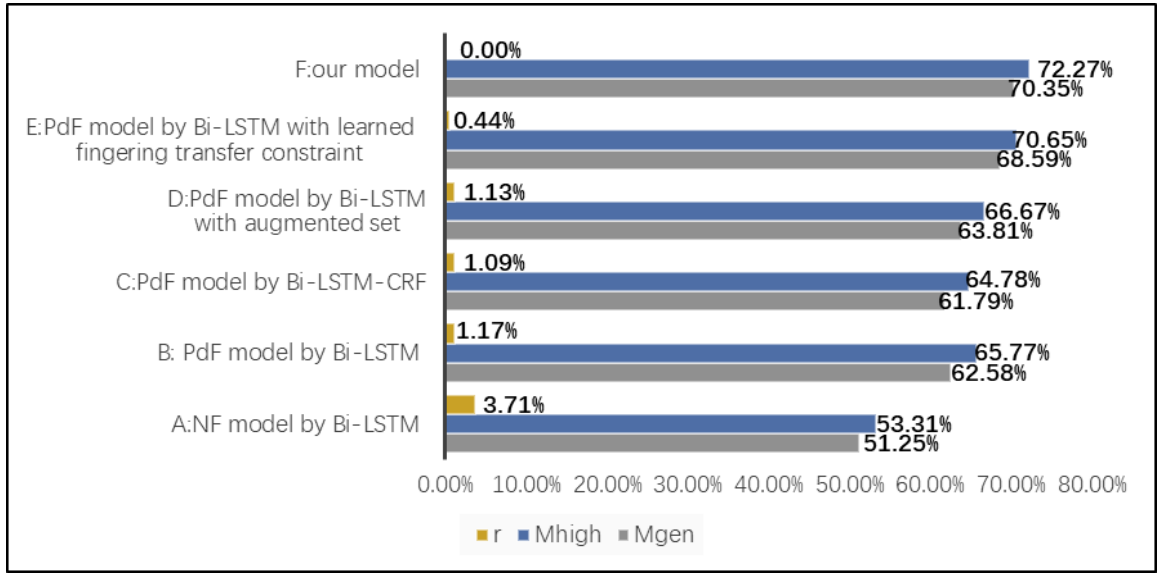

**Fig. 7** Results of ablation study

**4.3.4 Data extending**

The quantity and quality of training data is significant for deep learning-based models. Distilling the key factors of the training data while retaining as much information as possible is beneficial to reveal the mapping relationship between the original notes and fingering. On the other hand, generate an augmented set by analyzing original PIG dataset distribution is another effective strategy to alleviate model overfitting.

After adding the augmented set, the general and the highest matching rate are increased by 1.23% and 0.9% respectively, comparing result of B and D experiment in Fig. 7, which also confirms our belief that more data will help the system learn the difference information of fingering and pitch.

**4.3.5 Prior finger transfer knowledge**

Although we can see that the addition of fingering transfer knowledge greatly reduces the result irrationality, there are still some fingerings that will cause extra hand shifts or inability to play. As shown in experiment F in Fig. 7, addition of prior knowledge reduces IFR to 0. If there is a lot of data for training, the model may learn the correct fingering transfer knowledge by itself. At this time, the prior rules and the learned transfer knowledge can be combined into one, but at this stage, in order to ensure the playability of the annotation results, it is necessary to add prior constraint.

**4.3.6 Results of ablation study**

Strategy of modelling PdF relation, learned and prior finger transfer rule, and augmented dataset are all contributed to improve the performance of the fingering annotation system, among which, the new modelling strategy is the most significant factor which makes the annotaion accuracy increased by more than 10%, as shown in experimet A and B in Fig. 7. Based on this strategy, the refined model at the fingering constraint layer with learned finger tranfer regulation have further improved the model performance by about 5%, as shown in experimet B and E in Fig. 7. And by comparing their IFR,we can see that learnig of finger transfer regulation can effectively reduce the occurrence of unreasonble fingering. The addition of augmented set also can improve the performance, as shown in experimet B and D in Fig. 7. It also proves the effectivenesss of adding augmented set when data is limited.

**4.4 Example result and error analysis**

To directly demonstrate the concrete annotation result of models with different refining factor, the comparative fingering labels of different models under situations of long-range ascending scale, chord, and repeated notes are intercepted, as shown in Figure 8-10, respectively.

From annotated fingering of long-range ascending scale, right-hand score excerpt in Bach-Two-part invention in C major, in Fig. 8, where the marked red boxes show unplayable fingering, original BI-LSTM model in (b) gives continuous unplayable fingering among all methods. Compare annotated results in c and d, for learning relationship between fingers, learned fingering transfer constraint is better than CRF in this score excerpts, by which there are still two unplayable fingerings, however, when the pitch continues to rise, it is obviously much smoother with thumb under the finger than thumb shift in the green frame. The fingering results in (d) are not reasonable.

a: ground truth

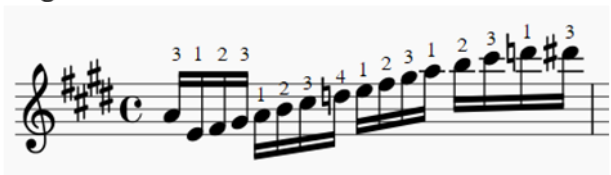

b: BI-LSTM model

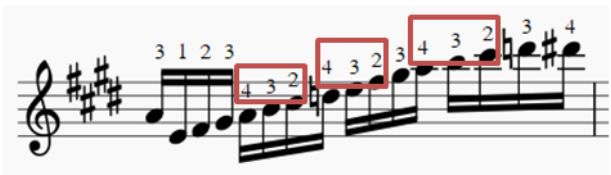

c: BI-LSTM CRF model

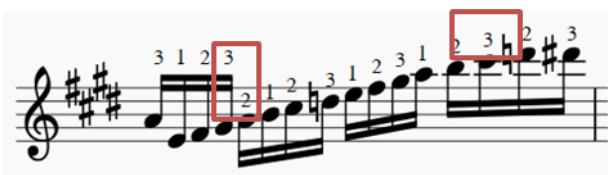

d: BI-LSTM with learned transfer constraint

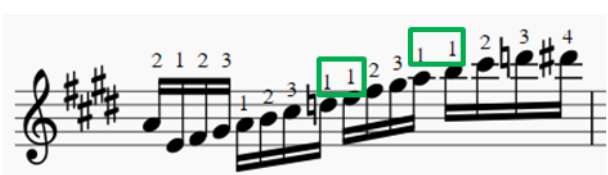

e: BI-LSTM with learned transfer constraint with augmented set

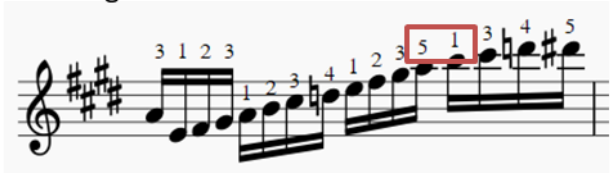

f: our model

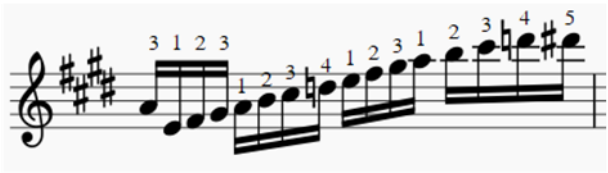

**Fig. 8** Annotation Comparison for long-range ascending scale

After adding the augmented set, as shown in Fig. 8(e), the two unreasonable fingerings decrease to one unplayable fingering from finger 5 to finger 1. After the fingering pruning was added in prediction period, the unplayable fingering was improved. Although the final annotation is still partly inconsistent with the ground truth, in terms of playability and comfortability, our result can be used in real performances. The result also shows that our model can improve the rationality of the result by adding the augmented set. And adding prior finger transfer knowledge can avoid unplayable fingering to a great extent and increase the flexibility of the result.

Fig. 9 shows the chord annotating results of the right-hand fragment of Scriabin Piano Sonata No. 5. Using note data without time information, as it shows in Fig. 9(b), it is easy to result fingerings that cannot be played. After preprocessing the note data into a pitch-difference with time in-formation, as it shows in Fig. 9(c), the unplayable fingerings decrease but has not disappeared.

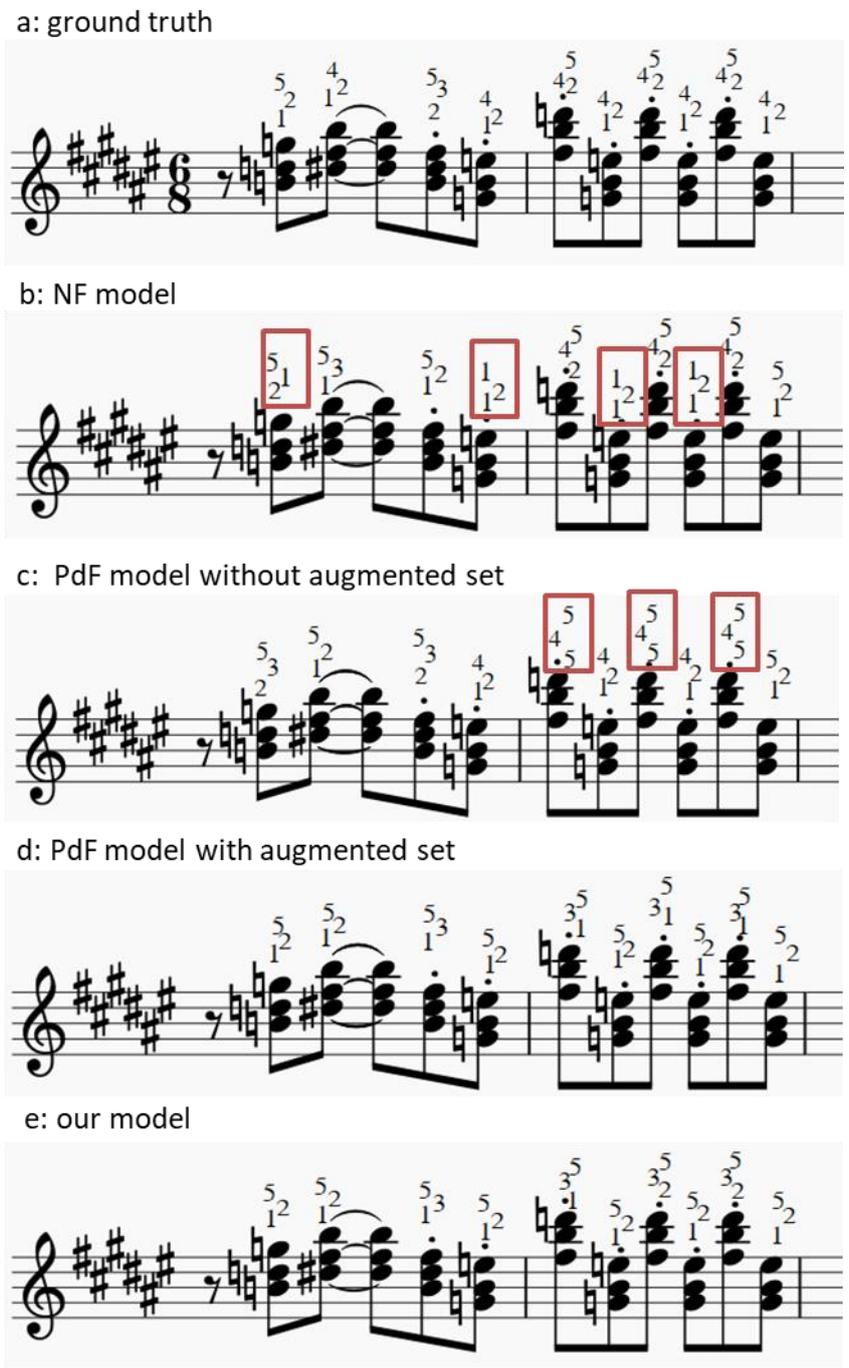

**Fig. 9** Annotation Comparison for chord

After comprehensive analysis, we found that the unplayable fingerings were concentrated on chords that did not appear in the training set. The elements of the augmented set include the correct fingering of the chord. It can be seen from Fig. 9(d) and Fig. 9(e), although adding the augmented set, it is still inconsistent with the ground truth, the playability of the fingering has been greatly improved.

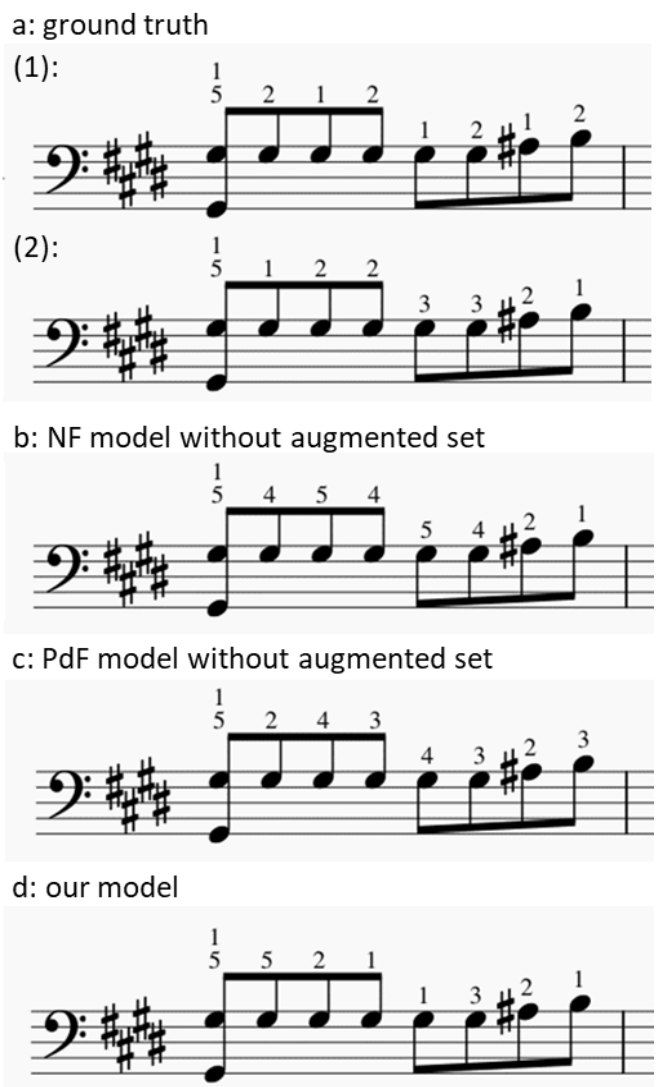

Fig. 10 Annotation Comparison for repeated notes

The annotation performance of left-hand score excerpt in Bach-Two-part invention in C major when the same tone appears repeatedly is given in Fig. 10. NF model and PdF model may have similar performance theoretically because of similar input type. But in case of a small amount of training data, PdF model can obviously classify more same tone into one category, that is, continuous '0'. The 4[th] finger appears in Fig. 10(b), (c). Due to the physiological structure of the hand, the fourth finger play the weakest among the 5 fingers, which is not suitable for playing the same pitch smoothly. Although no mandatory rules are added to avoid the use of 4th finger in the repeated fingering, 4th finger is replaced by other fingerings in our model. In addition, there is no continuous use of the same finger in each result, which shows that the long correlation of BI-LSTM can learn the corresponding features of music score and fingering Although there are still some differences between the model's result in Fig. 10(d) and the ground truths in Fig. 10(a), this result is more reasonable and can be played.

When there are many fingering combinations available, performers can choose fingerings completely according to their playing habits and comfort, and the algorithm has a larger choice of fingerings. Therefore, the use of matching rate as a single evaluation model will be biased.

**4.5 Accuracy comparison of different models**

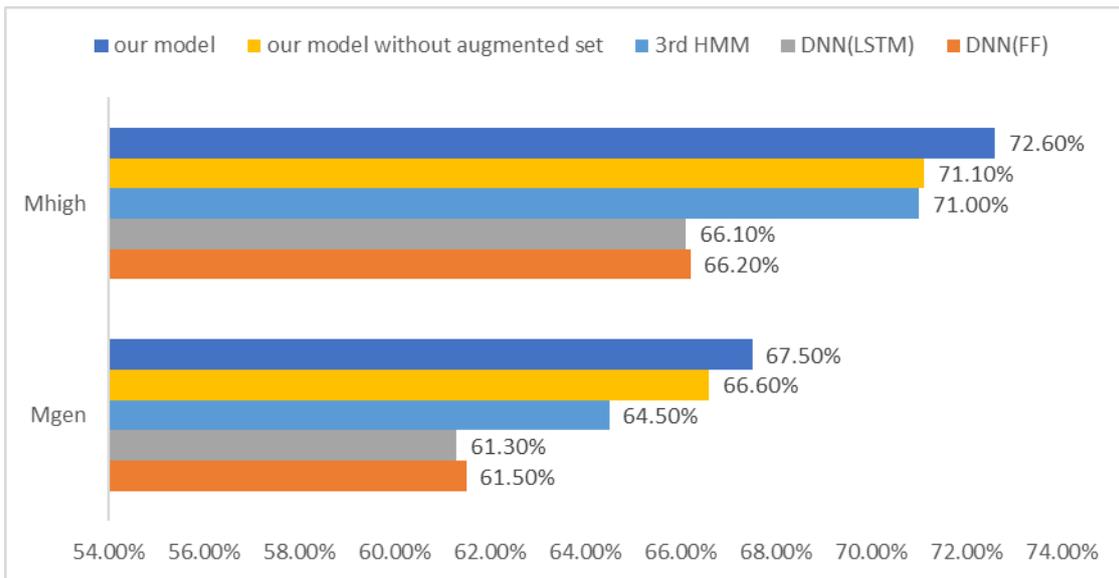

**Fig. 11** Comparison of labeling results of different algorithms

We compared our model with the state-of-the-art 3rd HMM, LSTM, and FF [9] in Fig. 11. The results show that our model outperform other methods. And for the best-performing third-order hidden Markov model, $M_{gen}$ has also increased by 3%, and $M_{high}$ has increased by 1.6%. Using the same training set, that is, without adding the augmented set, the model results have also improved by 2.1% and 0.1%.

## 5 Conclusion

We propose a playable piano fingering estimation method by pitch-difference fingering matching model with finger transfer knowledge which takes the pitch-difference sequence with time characteristics as input, combines the pitch context information through the output-constrained BI-LSTM neural network, introduce s the prior fingering rule in the recognition process, and chooses the fingering that conforms to the fingering rules with the highest performance efficiency. The result fingering is realistic and playable. Besides, we also generate an augmented set based on the existing data set PIG and fingering rules to alleviate overfitting problem caused by small training sets.

Our results show that pitch-difference with time and physical distance information is better suited for the task of fingering estimation than note information. Using BI-LSTM network and integrating fingering transfer constraints is more in line with the way of manually determining fingering. And the results verify that it is more helpful to consider the context rather than one-sided information to determine the fingering. Fingering constraints can ensure the playability of fingering. And when there are only 150 music score data at present, enlarging the data in a certain way or making full use of data, such as the addition of our augment set or using cross-validation, can indeed enable the model capture the characteristic relationship between pitch and fingering more effectively

The method proposed in this paper is better than the previous third-order hidden Markov method on the same test data according to matching rate. In terms of playability, our algorithm avoids unplayable fingerings, and all results are physically achievable. But our method also has some limitations. Here we summarize them.

(a) The performance of the model is improved by adding the constraint relationship between ad-

jacent fingerings $W_T$. The addition of long-range fingering transfer may make the model perform better.

(b) Because of the polymorphism and time transitivity of fingering, there is a certain degree of randomness in the manual and algorithmic determination of the labeling. Some scores give many ground truth fingering in order to overcome this property. However, the result combination of huge polymorphism is still somewhat lacking.

(c) When manually determining the fingering, there are many considerations such as the physical position and distance of the keys, the comfort of the hands, and the efficiency of performance. We have taken this factor into account in the algorithm, such as converting the pitch into a pitch-difference with time information, add fingering transfer, etc. But there is still some information, such as the cooperation of the hands is not considered in the model.

(d) The amount of data is insufficient. Although the augmented set is added, the information that the neural network can obtain is relatively limited. We believe that the algorithm will perform better when the data set is larger

Further investigations on these problems are left for future research.

**Appendix**

Generation method of augmented set

The generation of the augmented set is based on the fingering frequency distribution and the pitch-difference corresponding to the fingering in PIG[9]. Then, statistics are made on the pitch-difference with a frequency of more than 5% when all the scores before and after the fingering conversion. Among them, the polyphony is calculated as a whole, and some long context-related fingerings will be deleted from the data.

After obtaining the statistical information of the fingering, the transition probability matrix of the front and back fingerings is established. Taking into account the relatively simple rules and the transition probability, this generating model generates a total of 50 cases of data, and the length of each case is between 150 and 300. The augmented set generation principle is shown in Fig. A.1.

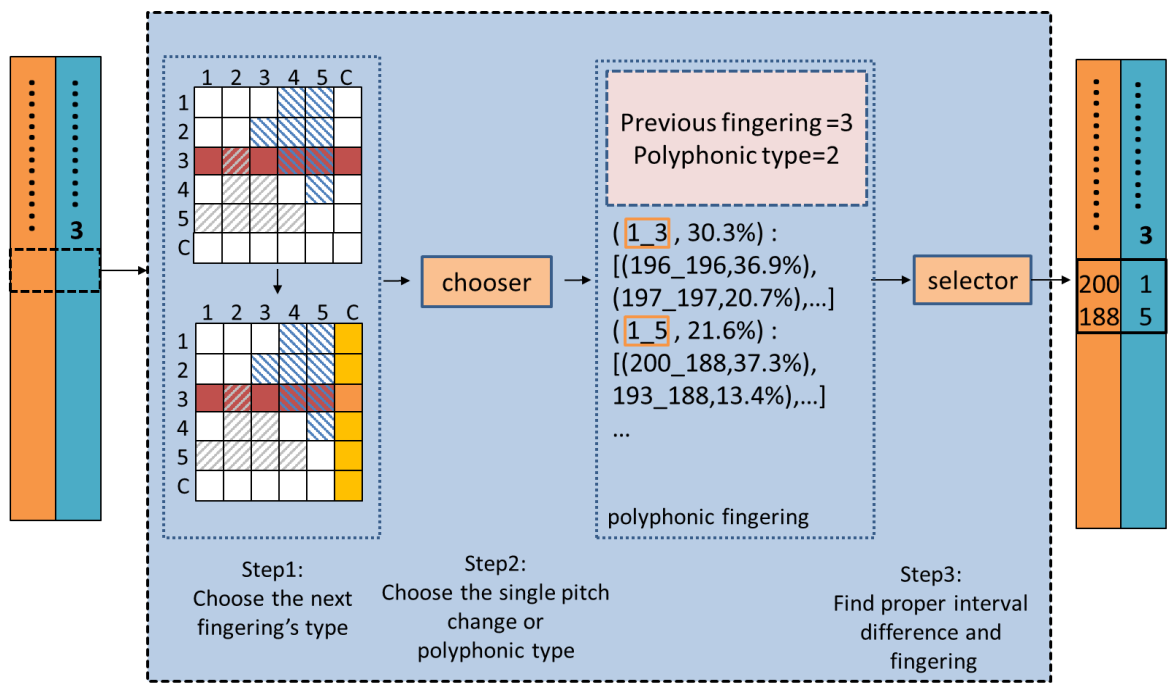

**Fig. A.1** Principle of augmented set generation

Fig. A.1 illustrates the generation process of the next fingering when the previous fingering is selected as $3^{th}$ fingers in the augmented set of left-hand fingering. The process is as follows: First, choose the type of the next fingering based on the probability. In the process of this Fig. A.1, the next fingering is selected as polyphony. In Fig. A.1, the 6*6 white grids represent the front and back fingering numbers. The gray shading indicates that the fingering conversion is not feasible at the point where the pitch of the single tone drops, and the blue indicates that the transition probability is 0 when the pitch is rising. If the single-tone fingering is selected, the next step is to select whether the corresponding pitch of the fingering is rising or falling according to the statistical information of the pitch rising or falling in the original data set; if the polyphonic fingering is selected, the type of polyphonic is selected as harmony It is still triad, seventh chord, etc. The chart selects the polyphonic type as harmony. Then according to the statistical probability, the selector selects the fingering information and the corresponding pitch-difference information to add to the extended set sequence. Repeat this process until the length of the fingering sequence reaches the specified length.